\documentclass{llncs}
\pdfoutput=1

\usepackage{graphicx}
\usepackage{amsmath,amssymb}
\let\doendproof\endproof
\renewcommand\endproof{~\hfill\qed\doendproof}

\begin{document}
\title{A note on planar partial 3-trees}
\author{Jan Kratochv\'\i l \and Michal Vaner}

\institute{
Institute for Theoretical Computer Science%
\thanks{Supported by the Ministry of Education of the Czech Republic as project 1M0545.}%
\\
and Department of Applied Mathematics,\\
Charles University, Prague, Czech Republic\\
\texttt{honza@kam.mff.cuni.cz, vorner@vorner.cz}
}

\maketitle

\begin{abstract}
It implicitly follows from the work of [Colbourn, El-Mallah: On two dual classes of planar graphs. Discrete Mathematics 80(1): 21-40 (1990)] that every planar partial 3-tree is a subgraph of a planar 3-tree. This fact has already enabled to prove a couple of results for planar partial 3-trees by induction on the structure of the underlying planar 3-tree completion. We provide an explicit proof of this observation and strengthen it by showing that
one can keep the plane drawing of the input graph unchanged.
\end{abstract}

\section{Overview of the result}

This paper is written with linguistic ambitions. Our aim is to explore the semantic difference in alternating the order of adjectives in an aggregated attribute construction. Specifically, we are interested in comparing the classes of {\em planar partial 3-trees} and {\em partial planar 3-trees}. But let us first argue why these classes should or should not coincide.

If ${\cal P}$ and ${\cal Q}$ are two (graph) properties then both expressions ${\cal P}{\cal Q}$-graphs and  ${\cal Q}{\cal P}$-graphs merely refer to {\em graphs that have both properties ${\cal P}$ and ${\cal Q}$}, i.e., to the class $\{G: G\in {\cal P}\cap {\cal Q}\}$ (as usual, we use $\cal P,Q$ both as names of properties and as names of the classes of graphs having these properties). As an example, the expression a {\em planar 3-colorable graph} has exactly the same meaning as a {\em 3-colorable planar graph}. The role of the adjective {\em partial} is, however, different. It literally means "take all subgraphs of". Thus {\em partial $\cal PQ$-graphs} are the subgraphs of graphs having both properties $\cal P$ and $\cal Q$, while {\em $\cal P$ partial $\cal Q$-graphs} are those subgraphs of graphs having property $\cal Q$ that also happen (the subgraphs) to have property $\cal P$. Formally

$$\mbox{"partial }{\cal PQ}\mbox{-graphs" } = \; \{G:\exists H\in {\cal P}\cap {\cal Q},\, G\subseteq H\}, \mbox{ and }$$
$$\mbox{"}{\cal P}\mbox{ partial }{\cal Q}\mbox{-graphs" } = \; \{G:\exists H\in {\cal Q},\, G\subseteq H, \mbox{ and } G\in {\cal P}\}. $$
If the property $\cal P$ is monotone, i.e., closed under taking subgraphs, then these two classes of graphs are in obvious inclusion, namely
{\em every "partial $\cal PQ$-graph" is a "$\cal P$ partial $\cal Q$-graph"},
but the converse is not necessarily true (e.g., all forests are acyclic subgraphs of cliques, while only graphs with at most 2 vertices are subgraphs of acyclic cliques). We hope that after this linguistic introduction the reader should pause in awe when exposed to the fact that for properties $\cal P$ = "being planar" and $\cal Q$ = "being a 3-tree", the two classes in question actually coincide.

But why are we interested in planar partial 3-trees? There is no need to explaining why we are interested in planar graphs in a paper devoted to graph drawing. Partial 3-trees are exactly graphs of tree-width at most 3, and as such important in the Robertson-Seymour theory of graph minors, as well as interesting from the computational complexity point of view (many decision and optimization problems that are hard for general graphs are polynomially solvable for graphs of bounded tree-width). Planar 3-trees are special types of planar triangulations (sometimes referred to as "stacked triangulations"), which can be generated from a triangle by a sequential addition of vertices of degree 3 inside (triangular) faces. As such they allow inductive proofs of many of their interesting properties (e.g., planar 3-trees are 4-list colorable, what is not true about all planar triangulations). Sometimes these proofs carry on for subgraphs, i.e., for partial planar 3-trees. In the area of graph drawing, we can list 3 examples:

\begin{itemize}
\item Badent et al. \cite{cccg} show that every planar partial  3-tree is a contact graph of homothetic triangles in the plane,
\item Biedl at Velazqez \cite{Biedl} show that every planar partial  3-tree has a straight line drawing in the plane with prescribed areas of the faces, and
\item Jel\'\i nkov\'a et al. \cite{slopes} show that planar partial  3-trees of bounded degree have straight line drawings in the plane with bounded number of slopes.
\end{itemize}
All these three results were proved by induction along the perfect elimination scheme for subgraphs of planar 3-trees, and the validity for the seemingly larger class of planar partial 3-trees follows  by the equivalence of these classes.

T. Biedl noted in a preliminary version of \cite{Biedl} that the equivalence follows implicitly from results of El-Mallah and Colbourn \cite{CM} (the relevant observation is made there in Corollary~4 which can be used to treat the case of 3-connected graphs). We provide here a detailed proof of a statement which is stronger in two aspects -- when augmenting a given planar partial 3-tree to a planar 3-tree it suffices to be adding edges only (unless the graph has less than 3 vertices), and if the input graph comes with a given noncrossing drawing in the plane, we can request this drawing in the augmented graph:

\begin{theorem}\label{thm-plane}
Every $n$-vertex planar partial 3-tree $G$ with $n\geq 3$ is a spanning subgraph
of a planar $n$-vertex 3-tree $\widetilde{G}$. Moreover, for any planar noncrossing drawing of $G$, the supergraph $\widetilde{G}$ can be constructed so that it has a planar noncrossing drawing that extends the one of $G$.
\end{theorem}

\section{Basic results on partial $k$-trees}

All considered graphs are undirected and without loops or multiple edges (we may allow multiple edges during some constructions, but delete them in the final steps). A graph is {\em chordal} if it does not contain an induced cycle of length greater than 3. A vertex is {\em simplicial} if its neighborhood induces a complete subgraph (i.e., a clique). It is well known that a graph is chordal if and only if it can reduced to the empty graph by sequential deletion of simplicial vertices. Equivalently, every chordal graphs allows a {\em perfect elimination scheme} (a PES, for short), which is an ordering $v_1, v_2, \ldots, v_n$ of its vertices such that each $v_i$ is simplicial in the subgraph induced by $v_1, v_2, \ldots, v_i$.

A chordal graph is a {\em $k$-tree} if it has a PES such that $v_1, v_2, \ldots, v_k$ induce a clique and each vertex $v_i, i>k$ has degree exactly $k$ in the subgraph induced by $v_1, v_2, \ldots, v_i$.
In such a case we say that the graph {\em is grown} from the starting clique $\{v_1, v_2, \ldots, v_k\}$.
A graph is a {\em partial $k$-tree} if it is a subgraph of a $k$-tree. The following two structural results hold for arbitrary $k$.

\begin{proposition}
{\rm \cite{Pro}} Every $k$-tree can be grown from any of its $k$-cliques.
\end{proposition}

\begin{proposition}
{\rm \cite{Pro}} Every partial $k$-tree with at least $k$ vertices is a spanning subgraph of a $k$-tree.
\end{proposition}

It is well known that for every fixed $k$, the class of partial $k$-trees is closed in the minor order, and hence can be characterized by a finite number of forbidden minors. Explicit characterizations are known only for small $k$'s. For our purposes the most interesting is the characterization of partial 3-trees by  four forbidden minors proven in \cite{AP}. This characterization is also used by El-Mallah and Colbourn in \cite{CM}. Our proof of Theorem~\ref{thm-plane} is straightforward and does not exploit it.

A 3-tree with $n$
vertices has $3+(n-3)3 = 3n-6$ edges, and hence every planar
3-tree is a planar triangulation, and as such it has a
topologically unique noncrossing embedding in the plane (up to the
choice of the outerface and its orientation). Such an embedding is
referred to as a {\em planar drawing} of the graph, or simply a
{\em plane graph}. As a consequence of the above propositions, a planar
3-tree can be grown from any of its triangles by consecutively
adding a vertex of degree 3 inside a face of the so far constructed
triangulation.

\section{Proof of Theorem~\ref{thm-plane}}\label{sec:plane}

\begin{proof} 
We prove the statement by induction on the number of vertices of
$G$. It is clearly true for $n=|V(G)|\le 4$. Hence we suppose that
$n>4$ and a plane graph $G$ on $n$ vertices is given.

If $G$ is disconnected, let $G$ be the disjoint union of nonempty
graphs $G_1$ and $G_2$ such that all vertices of $G_2$ lie in the
outerface of $G_1$, and all vertices of $G_1$ lie in the same face,
say $f$, of $G_2$, in the given planar drawing. Suppose for a moment
that both $G_1$ and $G_2$ have at least 3 vertices each. Both $G_1$
and $G_2$ are planar partial 3-trees, and hence by the induction
hypothesis, $G_i$ is a spanning subgraph of a planar 3-tree
$\widetilde{G_i}$ having a drawing which extends the drawing of $G_i$,
for $i=1,2$. Let $abc$ be the triangle whose edges bound the
outerface of $\widetilde{G_1}$ and let $xyz$ be a triangle in the
triangulation of $f$ in $\widetilde{G_2}$. We extend $G$ to
$\widetilde{G} = (V(G), E(\widetilde{G_1}) \cup E(\widetilde{G_2})
\cup \{xa,xb,xc, ya, yb, za\})$. This graph is planar and an
embedding which extends the embedding of $G$ is obtained from the drawing of
$\widetilde{G_2}$ by placing the drawing of $\widetilde{G_1}$ inside
the triangle $xyz$ and adding the edges $xa,xb,xc, ya, yb, za$. If
$P_1$ is any PES for $\widetilde{G_1}$, and $P_2 = xyz\ldots$ is a
PES for $\widetilde{G_2}$ (here we are using the fact that $\widetilde{G_2}$ can be grown from any of its triangles), the concatenation $P_1P_2$ is a PES for
$\widetilde{G}$, showing that $\widetilde{G}$ is a 3-tree. If one of
the graphs $G_1,G_2$ has at most 2 vertices, the other one has at
least 3 and by the induction hypothesis, it can be extended to a plane
3-tree. Then the vertex (-ices) of the first graph are added at the
tail of the PES. See an illustrative Figure~\ref{fig:disconnected}.

\begin{figure}[h]
\centering
\includegraphics[scale=0.35]{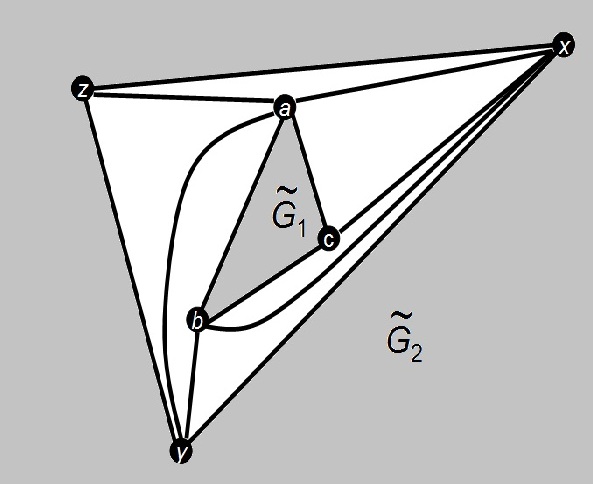}
\caption{Illustration to the disconnected case.} \label{fig:disconnected}
\end{figure}

Suppose $G$ is connected  but not 2-connected, and let $a$ be an
articulation vertex. Let $C$ be an inclusion-wise smallest connected
component of $G-a$ (in the sense of the given embedding). Consider
$G_1=G[C\cup\{a\}]$ and $G_2 = G[V(G)-C]$. Then all vertices of
$G_2$ (except of $a$) lie in the outerface of $G_1$, $a$ lies on the
boundary of the outerface of $G_1$, and all vertices of $G_1$
(except of $a$) lie inside one face, say $f$, of $G_2$, and this
face contains $a$ on its boundary. Suppose that both $G_1$ and $G_2$
have at least 3 vertices each (if one of them has at most 2
vertices, the argument is similar and even simpler).

By the induction hypothesis, $G_i$, for $i=1,2$, is a spanning subgraph
of a planar 3-tree $\widetilde{G_i}$ with a planar drawing which
extends the drawing of $G_i$. We may assume that $a$ lies on the
boundary of the outerface of $G_1$, since otherwise we can reroute
some of the added edges. Let $abc$ be the triangle whose edges bound
the outerface of $\widetilde{G_1}$ and let $axy$ be a triangle in
the triangulation of $f$ in $\widetilde{G_2}$ which contains $a$. We
extend $G$ to $\widetilde{G} = (V(G), E(\widetilde{G_1}) \cup
E(\widetilde{G_2}) \cup \{xb,xc, yc\})$. This graph is planar and an
embedding which extends the embedding of $G$ is obtained from
$\widetilde{G_2}$ by gluing the drawing of $\widetilde{G_1}$ inside the
triangle $axy$, unifying the vertices $a$ coming from
$\widetilde{G_1}$ and $\widetilde{G_2}$, and adding the edges
$xb,xc, yc$. If $P_1$ is any PES for $\widetilde{G_1}$, and $aP_2 =
axy\ldots$ is a PES for $\widetilde{G_2}$, the concatenation
$P_1P_2$ is a PES for $\widetilde{G}$ showing that $\widetilde{G}$
is a 3-tree.

\begin{figure}[h]
\centering
\includegraphics[scale=0.35]{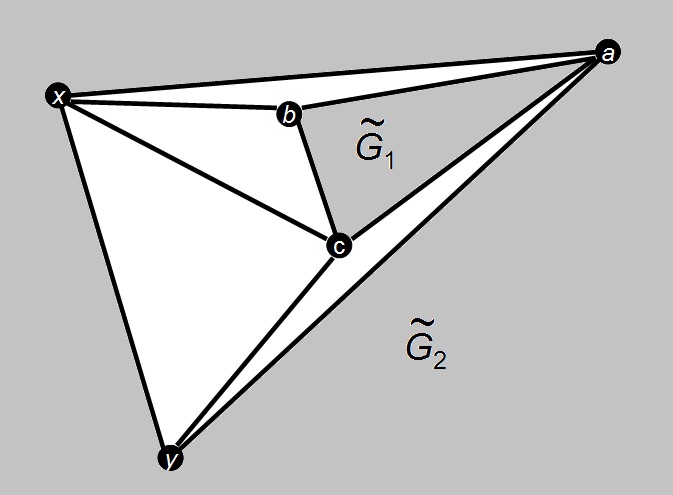}
\caption{Illustration to the simple connected case.} \label{fig:1connected}
\end{figure}

Suppose $G$ is 2-connected but not 3-connected. Let $a,b$ be a
minimal cut. Suppose that $G$ contains the edge $ab$. Let $C$ be an
inclusion-wise minimal connected component of $G[V(G)-\{a,b\}]$ such
that all vertices of $V(G)-(C\cup\{a,b\})$ lie in the outerface of
$G_1=G[C\cup\{a,b\}]$. Then the edge $ab$ belongs to the boundary of
this outerface and $C$ lies inside a face, say $f$, of
$G_2=C[V(G)-C]$. The edge $ab$ belongs to the boundary of $f$, too.

By the induction hypothesis, $G_i$, for $i=1,2$, is a spanning subgraph
of a planar 3-tree $\widetilde{G_i}$ with a planar drawing which
extends the drawing of $G_i$. We may assume that $ab$ lies on the
boundary of the outerface of $G_1$, since otherwise we can reroute
some of the added edges. Let $abc$ be the triangle whose edges bound
the outerface of $\widetilde{G_1}$ and let $abx$ be a triangle in
the triangulation of $f$ in $\widetilde{G_2}$ which contains $ab$.
We extend $G$ to $\widetilde{G} = (V(G), E(\widetilde{G_1}) \cup
E(\widetilde{G_2}) \cup \{xc\})$. This graph is planar and an
embedding which extends the embedding of $G$ is obtained from
$\widetilde{G_2}$ by gluing the drawing $\widetilde{G_1}$ inside the
triangle $axy$, unifying the edges $ab$ coming from
$\widetilde{G_1}$ and $\widetilde{G_2}$, and adding the edge $xc$.
If $P_1$ is any PES for $\widetilde{G_1}$, and $abP_2 = abx\ldots$
is a PES for $\widetilde{G_2}$, then the concatenation $P_1P_2$ is a
PES for $\widetilde{G}$ showing that $\widetilde{G}$ is a 3-tree.

If $ab$ is not an edge of $G$, we can add it arbitrarily in the
drawing of $G$. Straightforwardly, the graph remains planar. Then
$G_1+ab$ (as well as $G_2+ab$ are planar partial 3-trees and by the
induction hypothesis can be extended to planar 3-trees, which can be
glued together along the edge $ab$. Therefore $\widetilde{G}$
(constructed as in the previous case) is a planar 3-tree.

\begin{figure}[h]
\centering
\includegraphics[scale=0.4]{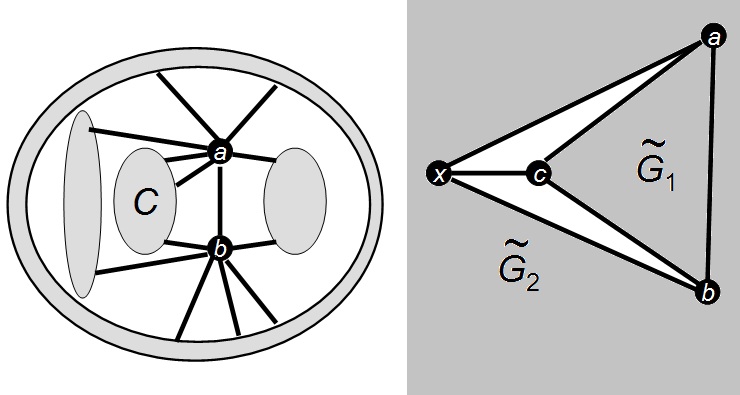}
\caption{Illustration to the 2-connected case.} \label{fig:2connected}
\end{figure}

Suppose $G$ is 3-connected, in particular, every vertex of $G$ has degree at least 3.
By assumption $G$ is a partial 3-tree, and hence some 3-tree $G'$ is a supergraph of $G$.
Let $u_1,\ldots,u_h$ be a PES for $G'$ (note we do not assume $h=n$ at this point).
If $G'$ is a vertex minimal such supergraph, we conclude that $u_h\in V(G)$ (otherwise we could just
delete this vertex from $G'$). Let $\{a,b,c\}$ be the neighbors of $u_h$ in $G'$. Since $G$ is 3-connected, all three vertices $a,b,c$ belong to $G$. Since $u_h$ is simplicial in $G'$, all three edges $ab,bc,ac$ belong to $G'$ and hence $\overline{G}=(V(G),E(G)\cup\{ab,ac,bc\})$ is a partial 3-tree. We also conclude that $\overline{G}$ is planar, since the added edges can be drawn in the angles $au_hb, bu_hc, cu_ha$ without crossing other edges of $G$.

Since $G$ is 3-connected, $a,b,c$ is a minimal cut and $G[V(G)-\{a,b,c\}]$ has exactly 2 connected components, one of them being the vertex $u_h$ on its own (otherwise $G$ would contain $K_{3,3}$ as a minor and would not be planar).
By the induction hypothesis, $\overline{G}-u_h$ is a spanning subgraph of a planar 3-tree which has a drawing extending the one of $G-u_h$. In this drawing the triangle $abc$ induces an empty face (since $G[V(G)-\{a,b,c,u_h\}]$ is connected). Thus we can embed the vertex $u_h$ in the face $abc$ of this drawing.
\end{proof}

\section{Acknowledgment}

We thank Robin Thomas for valuable discussions leading to the proof presented in Section~\ref{sec:plane}.


\begin{thebibliography}{ccc}

\bibitem{AP}
Stefan Arnborg, Andrzej Proskurowski, Derek G. Corneil: Forbidden minors characterization of partial 3-trees. Discrete Mathematics 80(1): 1-19 (1990)

\bibitem{cccg}
Melanie Badent, Carla Binucci, Emilio Di Giacomo, Walter Didimo, Stefan Felsner, Francesco Giordano, Jan Kratochv\'\i l, Pietro Palladino, Maurizio Patrignani, Francesco Trotta: Homothetic Triangle Contact Representations of Planar Graphs. CCCG 2007: 233-236

\bibitem{Biedl}
Therese C. Biedl, Lesvia Elena Ruiz Vel\'azquez: Drawing Planar 3-Trees with Given Face-Areas. in: Graph Drawing 2009,
LNCS 5849, Springer 2010, pp. 316-322

\bibitem{CM}
Ehab S. El-Mallah, Charles J. Colbourn: On two dual classes of planar graphs. Discrete Mathematics 80(1): 21-40 (1990)

\bibitem{slopes}
Vit Jel\'\i nek, Eva Jel\'\i nkov\'a, Jan Kratochv\'\i l, Bernard Lidick\'y, Marek Tesa\v r, Tom\'a\v s Vysko\v cil: The Planar Slope Number of Planar Partial 3-Trees of Bounded Degree. in: Graph Drawing 2009, LNCS 5849, Springer 2010, pp. 304-315


\bibitem{Pro}
Andrzej Proskurowski: Separating subgraphs in k-trees: Cables and caterpillars. Discrete Mathematics 49(3): 275-285 (1984)

\end{thebibliography}
\end{document}